# Structural and electronic properties of $V_2O_5$ and $MoO_3$ bulk and ultrathin layers

Tilak Das, Sergio Tosoni, and Gianfranco Pacchioni*

*Dipartimento di Scienza dei Materiali, Universita' degli Studi Milano-Bicocca,
via R. Cozzi, 55 - 20125 Milano, Italy*

Version 10.12.2018

*Corresponding Author's Email: gianfranco.pacchioni@unimib.it

**ABSTRACT**: The structural and electronic properties of bulk, monolayer and ultrathin films of $V_2O_5$ and $MoO_3$ layered oxides have been studied with first-principles density functional theory calculations including Van der Waals dispersion corrections. Specific DFT+*U* functionals have been tested in order to properly reproduce geometry, band-gap, static dielectric constant, and formational enthalpies of the two materials. The mono-, and multi-layers are cleaved along the <001> and <010> stable crystallographic orientations for $V_2O_5$ and $MoO_3$, respectively. At least three layers are needed for both materials in order to recover bulk-like properties. Spin-orbit effects have been incorporated in our prediction, but they show marginal effects.

1. **Introduction**

In recent years an increasing interest has been devoted to low dimensional materials, and in particular to two-dimensional, or 2D materials.[1,2] In many cases these systems can be exfoliated separating the single layers from the rest of the material. Successful examples have been reported for transition metal dichalcogenides (TMDs) beside the well known graphene case.[3,4] For some three-dimensional oxides it is also possible to growth ultrathin films (2D oxides) on a metal support; ultrathin films of MgO, $Al_2O_3$, FeO, $SiO_2$, ZnO etc. have been grown on various metal substrates. The properties of these systems are often quite different from those of the corresponding bulk materials.[5] This is also due to the fact that often 2D oxides assume structures that are markedly different from the corresponding bulk structures. On the other hand, layered oxides like vanadium pentoxide ($V_2O_5$) or molybdenum trioxide ($MoO_3$) exist in nature, and exfoliation of these layered oxides can in principle be used to obtain new materials with novel properties and applications.[6,7] For instance, a recent study has shown the potential of liquid exfoliation of $MoO_3$ nanosheets to obtain new materials for supercapacitors energy storage applications.[8]

The thermodynamically stable orthorhombic phase (also known as α-phase) of $V_2O_5$ and $MoO_3$, two transition metal oxides with $d^0$ character, have been widely studied both experimentally and theoretically due to their excellent opto-electronic, photo- and electro-chemical properties, also useful for catalysis.[9,10] Recently, in the search of metal-ion-battery electrodes, $V_2O_5$ has been proposed as a good candidate for Li-ion batteries with theoretical capacity up to 294 mA h $g^{-1}$.[11,12] On the other hand, $MoO_3$ is also well known for technological applications as an inorganic photochromic and electrochromic material,[13,14] as catalyst,[15,16] and sensor.[17] The thermodynamically stable surface is (001) for $V_2O_5$,[18] and (010) for $MoO_3$.[6] Using a liquid exfoliation technique Rui et al. reported the preparation of $V_2O_5$ 2D nanosheets of width 2.1 to 3.8 nm.[11] Other routes to obtain $V_2O_5$ nanostructures from hydrothermal, solvothermal, and wet/dry exfoliation techniques have been reported.[12] As for $MoO_3$, Molina-Mandoza et al.[6] have studied the dependence of the properties on the thickness but no major change in band-gap and band-structure has been observed going from the monolayer to the bulk. This result contrasts with a recent study by Liu et al.[7] based on the preparation of micro- to nano-sized crystals of $MoO_3$. According to this study, the optical band-gap shows a clear blue shift (~1.5 eV) going from 4 to 8 layers, and to the bulk phase of $MoO_3$.

There are structural similarities in these two oxides, and a similar arrangement via inter-layer Van der Waals (vdW) interactions. The main difference is in the metal-oxygen polyhedral building blocks and their chain structure. In $V_2O_5$ each V atom is connected to five oxygen atoms to form a square pyramid; in $MoO_3$ each Mo atom is connected to six oxygen atoms in octahedral coordination. These pyramids and octahedral building blocks are connected via edge sharing oxygen atoms and form elongated chains along the *xy*-crystallographic (*xz*-) plane, separated by a Van der Waals spacing.

Both $V_2O_5$ and $MoO_3$ bulk crystals have three types of oxygen atoms, namely an apex vanadyl or molybdenyl oxygen, $O_v$ and $O_m$, a chain oxygen, $O_c$, and a bridging oxygen between the chains, $O_b$. The minimum inter-layer spacing (*d*) between layers in the bulk structure is wave-like (zigzag) due to the oxygen atoms arrangement, and is slightly lower for $V_2O_5$ (2.321 Å) than for $MoO_3$ (2.644 Å).

Alongside with the large number of experimental studies of electronic and optical properties of $V_2O_5$ and $MoO_3$, either in bulk or thin-films geometry, a number of DFT studies has been reported. The PW91+*U* GGA functional has been used to study the impact of different values of the *U* parameter on V(*d*) states on the geometry and electron localization.[18] In another study, GW calculations (over the LDA ground state)[19] have been reported on a $V_2O_5$ monolayer, with a predicted band gap of 4 eV. Moving to $MoO_3$, PBE-D2 DFT calculations have been performed introducing compressive and tensile strain to explain some experimental findings.[7] In a study of bulk, monolayer and nano-ribbons of $MoO_3$ (PBE-D2+*U* DFT functional), the authors have reported a tiny change of the indirect band-gap of bulk (1.71 eV) compared to monolayer (1.73 eV).[20] Lutfalla et al. have discussed the formation enthalpy from the possible redox pairs of $V_2O_5/VO_2/V_2O_3$ and $MoO_3/MoO_2$.[27] From their *ad hoc* choice of *U* value from 0-10 eV, the predicted effective *U* value for V(*d*) and Mo(*d*) are respectively ~3 eV and ~8 eV, which gives the best thermochemistry for the chemical reduction of $V_2O_5$ and $MoO_3$ bulk phase, respectively. Finally, Inzani et al. 2016 have checked many possible Van der Waals density functionals (vdW-DF), along with *U* correction on Mo(*d*). They have concluded that the geometry was better described with a vdW-D2 functional, and that there is almost no effect on the band-gap varying *U* on Mo(*d*) from 2 to 8 eV.[21]

The scope of this paper is twofold. On one side we will try to define a computational setup which is able to describe with a similar accuracy not only the structural parameters and the band gap of the material, which are the usual targets of the DFT calculations, but also of the dielectric constant and of the formation energy of the bulk material. This in view of the possible extension of the investigation of these materials in chemical processes where thermodynamic aspects are important. The second objective is to investigate to what extent the properties change by going from the monolayer to the bulk of $V_2O_5$ and $MoO_3$.

## 2. Choice of computational methodology

It is well known that in DFT studies of molecules and solids the choice of the exchange-correlation functional is critical. In particular, semiconducting and insulating materials need to use the so-called self-interaction corrected functionals, where the problem of the self-interaction error inherent in LDA and GGA approaches is partly removed. The problem is particularly severe for the strongly correlated *d*-orbitals of transition metal oxides.[22] To this end, two practical approaches are usually followed, one based on hybrid functionals, where a portion of the exact Fock exchange is mixed-in with the DFT exchange, and the other one based on the "on-site" Hubbard *U* correction for the *correlated* electrons. In both cases, the extent of exchange mixing, $\alpha$,[23,24,25] and the value of the *U* term[26,27] are determined in a more or less empirical way, by choosing the values that lead to a better reproduction of some key quantities (e.g. the lattice constants, the band gap of the material, etc.). An alternative which has been recently suggested consists in using a self-consistent hybrid approach where the amount of exact exchange $\alpha$ is not used as an external parameter but is varied in order to determine the static dielectric constant of the material. This approach is based on the observation that the static dielectric constant of a material $\varepsilon$ depends on $\alpha$, $\varepsilon = 1/\alpha$. Starting from an initial value of $\alpha$, one can compute $\varepsilon$ which is then used to determine a new value of $\alpha$ until the process leads to a stable, self-consistent $\varepsilon$ value. The method is called dielectric-dependent self-consistent hybrid approach.[28]

In this work we have both the dielectric-dependent hybrid functional and the DFT+*U* approach to study $V_2O_5$ and $MoO_3$ layered materials. We started by using the Perdew-Burke-Ernzerhof (PBE-GGA) formulation[29] of GGA as implemented in the plane-wave pseudopotential code Vienna Ab-initio Simulation Package (VASP).[30,31] The DFT+*U* calculations have been performed following the approach suggested by Dudarev et al.[22] The valence electrons are approximated with Projector Augmented Wave (PAW) method[32] with a plane-wave cut-off of 600 eV. A total of 13 valence electrons for V($3s^23p^64s^23d^3$) and 14 electrons for Mo($4s^24p^65s^24d^4$) have been considered explicitly; the O atom valence configuration is O($2s^22p^4$). All calculations were done using spin-polarization.

In order to describe the inter-layer interaction dominated by vdW forces, we have considered the long-range dispersion energy correction in the semi-empirical approach proposed by Grimme (3$^{rd}$ generation of dispersion force correction, DFT-D3).[33,34] The method is thus referred to as PBE+U/D3. Spin-orbit coupling (SOC) has also been considered in selected cases, via perturbation theory using a scalar-relativistic wave-function of the valence states in PBE+U/D3/SOC.

A convergence of the total energy and Hellmann-Feynman forces on the atoms of $1\times10^{-8}$ and 3 meV/Å, respectively, were adopted, using a Monkhorst-Pack k-mesh grid 4×6×5 (5×4×6) for $V_2O_5$ ($MoO_3$). For the mono or few-layers films the supercell models include a vacuum region of about 20 Å.

For the hybrid functional calculations we used the Crystal17 code based on localized basis functions.[35] We used the B3LYP[23,24,25] formulation of the hybrid functional, either in the original formulation where α = 0.20, or taking for α the inverse of the experimental dielectric constant for both $MoO_3$ and $V_2O_5$ bulk phases. All electron Gaussian-type basis-set 8-411G(d1) was used for O atom and 86-411G(d3) was used for V atom. For the heavy atom Mo, we have adopted a small core effective core potential (ECP), leaving the 4s, 4p, 4d, 5s states in the valence (14 valence electrons and basis-set).[36] The convergence of the total energy of the two consecutive SCF cycles were set to $1\times10^{-8}$ using the Monkhorst-Pack/Gilat shrinking factor 8/16 for bulk and slab models. The tolerance for the Coulomb overlap, exchange-overlap and penetration integral in direct space were set to $1\times10^{-7}$, whereas the default for the reciprocal space Coulomb exchange-pseudo overlap integral was set to $1\times10^{-14}$ in all calculations using Crystal17 code.

The scope of the first part of the work is to try to determine at a similar level of accuracy, not only the lattice parameters and the band gap of the two materials, but also their formation enthalpies and dielectric constants. Furthermore, the formation of an oxygen vacancy and the degree of electron localization have also been used as a test of the validity of the computational approach used. $V_2O_5$ and $MoO_3$ are two oxides widely used in catalysis, where their chemical activity does not depends only on the geometric structure and electronic band gap, but also on the cost of removing oxygen from the surface, e.g. in oxidative chemical processes. In the following we will discuss the results obtained, in order to arrive to a good compromise in terms of accuracy and computational cost.

## *2.1 Results of dielectric dependent hybrid calculations*

In this section we will discuss the results of hybrid functional calculations using the CRYSTAL17 code. These calculations have only an exploratory purpose and, as we will show below, are reported here in order to show that the use of a dielectric-dependent functional does not lead to a good description of the material. For the calculations we adopted the B3LYP hybrid functional, but in case of $MoO_3$ also the PBE0 functional has been considered. The calculations have been performed using a fixed optimal geometry for $V_2O_5$ and $MoO_3$ obtained at the PBE+U/D3/SOC level of theory. This structure is rather close to the experimental one, and the results are not expected to depend on small changes in the lattice parameters related to the use of an hybrid functional.

A self-consistent hybrid functional has been derived; the exchange fraction α was evaluated self-consistently using the computed average dielectric constant, ε, of the bulk oxides. For $MoO_3$ the computed value of this quantity, 4.26, differs by more than 30% from the experimental value, 5.7; the corresponding exchange fraction, given by the inverse of the dielectric constant, is 23.45%. With this value of α the Kohn-Sham band gap of the material is predicted to be 3.5 eV, while the experimental value is of about 2.2 eV. We repeated the procedure using the PBE0 formulation of the hybrid functional, but the results are similar: ε is 4.28 and the corresponding band gap 3.49 eV, far from the experimental values.

Next, we considered $V_2O_5$ using the same procedure: the results are qualitatively similar, with the DFT calculations predicting an average dielectric constant, 4.46, which differs substantially from the measured one, 5.9. We also checked that this is not related to the anisotropy of the computed dielectric constant. The result is that a band gap of 3.35 eV is predicted for $V_2O_5$, ~1 eV larger than in the experiment (2.3 eV)[37].

These data show that the use of a dielectric-dependent hybrid functional does not provides a good description of the band gap and of the dielectric properties of the two oxides. Further work is planned to better understand the reasons for this failure. For the rest of the study we decided to consider the PBE+U/D3 approach and to define a set of parameters able to describe at an acceptable level of accuracy a series of electronic and geometric properties. The rest of the discussion is based on this kind of approach. Most of the details about the determination of the set of parameters are reported in the Supplementary Information, Section S1, S2 and S3.

## 3. Properties of bulk $V_2O_5$

As we mentioned before, the final value of $U$ used in the calculations is the result of the analysis of four different properties, trying to provide a balanced and uniform description of structural, electronic, dielectric and thermodynamic properties. The effective $U$ value that seems to give this kind of balance for $V_2O_5$ is $U_V$ = 3.5 eV for V(*d*) within the PBE+U/D3/SOC approach. In the following we have provided evidence in support of this choice.

### *3.1 Structural parameters*

Bulk $V_2O_5$ belongs to the orthorhombic space group *P mmn*. The experimental lattice parameters are *a* = 11.512 Å, *b* = 3.564 Å, *c* = 4.368 Å.[38] The $V_2O_5$ bulk phase structure is shown in **Figure 1**. In panel [a] the unit cell is marked with a black solid line, and the optimal angles are marked; in panel [b], the $VO_5$ square pyramids are shown. They form a chain along the *y*-axis, separated by a wave like vdW inter-layer spacing marked with *d*. Each of these slabs is formed by $VO_5$ units and constitutes a monolayer of $V_2O_5$.

The calculated structural parameters and band-gap values can be found in **Table 1,** where they are compared to the experimental data for $V_2O_5$. The dispersion energy correction via vdW-D3 formalism is quite crucial, since the vdW interlayer spacing is reduced by 0.3 Å compared to a standard PBE approach. In general, the lattice parameters are well described, with errors of about 1%, at the PBE+U/D3/SOC level.

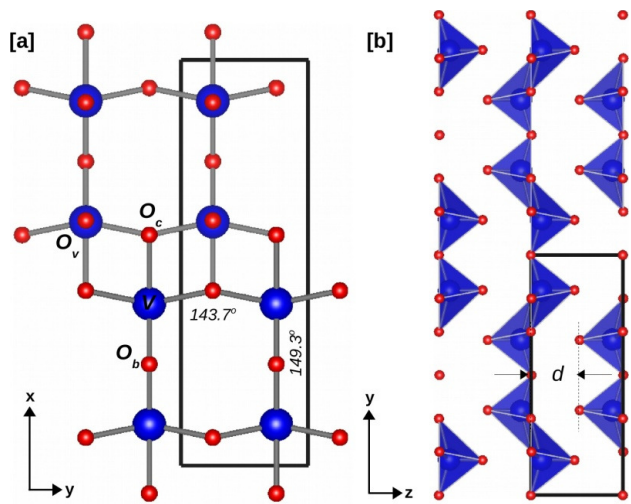

**Figure 1**: Optimized bulk crystal structure of α-V$_2$O$_5$ (PBE+U/D3/SOC functional with $U_V$ = 3.5 eV).

**Table 1**: Optimized bulk structure parameters of V$_2$O$_5$ from PBE and PBE+U/D3/SOC functionals ($U_V$ = 3.5 eV) compared with experimental data. In the parentheses is reported the % error with respect to experiment.

|  | Orthorhombic α-V$_2$O$_5$ (*P mmn*) | | |
| --- | --- | --- | --- |
| Parameters | PBE | PBE+U/D3/SOC | Exp.[38,43] |
| a(Å) | 11.551 | 11.628 (+1.0%) | 11.512 |
| b(Å) | 3.567 | 3.615 (+1.4%) | 3.564 |
| c(Å) | 4.716 | 4.359 (-0.3%) | 4.368 |
| vdW Spacing, *d* (Å) | 2.637 | 2.280 (-1.8%) | 2.321 |
| V-O$_v$ (Å) | 1.591 | 1.602 | 1.585 |
| V-O$_b$ (Å) | 2×1.789 | 2×1.798 | 2×1.780 |
| V-O$_c$ (Å) | 2×1.889 | 2×1.902 | 2×1.878 |
|  | 1×2.044 | 1×2.034 | 1×2.021 |
| <V-O$_b$-V (°) | 147.5° | 149.3° (+0.7%) | 148.2° |
| <V-O$_c$-V (°) | 141.7° | 143.7° (+0.3%) | 143.2° |
| Band-gap (eV) | 1.98 | 2.18 (-5.5%) | 2.3 |

### *3.2 Band gap and electronic structure*

V$_2$O$_5$ is a semiconducting oxide with a indirect band gap of 2.3 eV.[37] In the literature, various DFT+*U* studies have been reported, with the effective *U* values for the V(3*d*) states ranging from 2.3 to 6.6 eV, usually defined so as to reproduce the experimental band-gap. Recently, a quasi-particle self-consistent *GW* calculation predicted a band gap of 4 eV, much larger than the experimental one. It should be mentioned that the discussion here is based on the comparison of the Kohn-Sham[39] band gap with experiments. Most experimental band gaps are obtained from optical measurements, and due to excitonic effects optical band gaps are generally smaller than electronic band gaps. In transition metal oxides the exciton binding energy can vary from a few meV to several tenth of an eV.[40,41] In this respect, comparing Kohn-Sham band gaps with optical band gaps is a crude approximation, and one should not expect a one-to-one correspondence between the two values.

In order to obtain a small Kohn-Sham band gap in $V_2O_5$ relatively large values of the $U$ term have been used. $U = 6.6$ eV was adopted by Laubach et al.[42], but this turned out to be inappropriate as demonstrated later by Scanlon et al.[43] In fact, these authors used various $U$ values for $V_2O_5$, reporting a value of 2.30-2.35 eV for the band gap, in good agreement with the experiment. It should be noted, however, that in this study the $c$ lattice parameter was overestimated by nearly 10% due to the lack of dispersion corrections. In the study by Jovanović et al.[44] $U= 6$ eV was adopted to study transition metal ion doping in $V_2O_5$. Once more, while large $U$ values can provide a reasonable Kohn-Sham band gap, they may also have counter-effects in terms of defect states in the band gap and of thermochemistry, resulting in inaccurate properties. In this context, a smaller value $U = 3.1$ eV has been prescribed to study the intercalation of Mg ions in different phases of $V_2O_5$.[45]

In our study, using $U = 3.5$ eV, the band gap of the material is quite well reproduced, 2.18 eV against 2.3 eV (Table 1). However, we notice that the inclusion of the $U$ term has only a moderate effect on this property which goes from 1.98 eV at PBE level, to 2.18 eV after inclusion of vdW forces, $U$ term and SOC contributions.

In **Figure 2** we report the calculated density of states, DOS (left panel) and the total band structure (right panel) of bulk pristine $V_2O_5$. The indirect transition from R-to-Γ k-path of the high symmetry k-points of the first Brillouin zone of bulk $V_2O_5$ leads to opening of the lowest fundamental band-gap. In the top panel of DOS, the total and V($d$) atom projected DOS is shown. In the lower panel, the total O($p$) DOS for all three types of oxygen atoms are shown, where a clear difference is seen for $O_c$ (or $O_b$) compared to the $O_v$ type oxygen atoms.

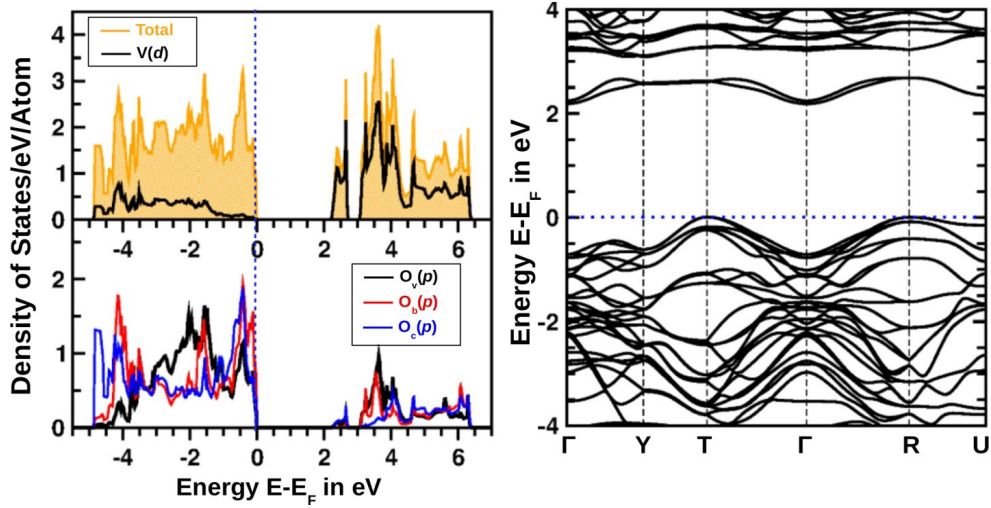

**Figure 2**: Total, V(*d*), and O(*p*) projected density of states (left panel) and band-structure (right panel) of bulk $V_2O_5$ (PBE+U/D3/SOC). The Fermi level is set to the top of the valence band.

From these plots it is clear that the top of the valence band is essentially formed by the O(*p*) states and bottom of the conduction band is formed from a splitted V(*d*) band, which results in a reduced band-gap for the $V_2O_5$ layered oxide. Details of the projected DOS of V(*d*) are given in the SI **Section S4, Figure 1**.

The impact of SOC has been checked, but this term does not alter the band-splitting (the band-edges of the valence and conduction bands move by less than 2 meV). Thus, spin-orbit coupling will be no longer discussed.

*3.3 Formation Enthalpy*

The formation enthalpy (*ΔH*) of the bulk phase of vanadium pentoxide has been considered starting from solid vanadium metal and molecular oxygen. Experimentally, the formation of $V_2O_5$ releases 1550.6 KJ/mol (16.15 eV/f.u.). At the PBE+U/D3/SOC level ($U = 3.5$ eV), the computed formation enthalpy is 15.93 eV, i.e. remarkably close to the experimental value (error of 1.4%) (see **Section S2** for details).

*3.4 Static Dielectric Constant*

The last property considered is the static dielectric constant, ε, calculated using the Kramers-Kronig dispersion relations within linear optical properties and independent particle approach, as implemented in the VASP code.[46] Details of the optical equations, frequency dependent real and imaginary dielectric tensor are given in the SI **Section S3**. The calculated values of ε along all three polarization axis of $V_2O_5$ are tabulated in **Table 2** and compared with the experimental data. Note that in this table the only trace component of the real part of the dielectric constant as computed at zero frequency along the three principle crystallographic axis are shown, and there average. We did not find previous DFT studies of the static dielectric constant for $V_2O_5$. Our prediction, ε = 5.65, reproduces quite satisfactorily the known experimental value 5.9, with an error of 4.2%.[47] Notice that some effect is found in the calculation of this property by going from the standard PBE functional to the PBE+U/D3/SOC ne. In particular, there is a stronger anisotropy of the constant at the higher level of theory, which probably reflects the different interlayer distance. The final average value, however, is not particularly affected by the choice of the method.

**Table 2**: Calculated static dielectric constant, ε, of bulk $V_2O_5$ from PBE and PBE+U/D3/SOC approaches compared with experiment.

| | Orthorhombic α-$V_2O_5$ (*P mmn*) | | | |
|---|---|---|---|---|
| Functionals | $\varepsilon_{1xx}(0)$ | $\varepsilon_{1yy}(0)$ | $\varepsilon_{1zz}(0)$ | $\varepsilon_{avg}(0)$ |
| PBE | 6.34 | 5.83 | 5.87 | 6.01 |
| PBE+U/D3/SOC | 5.88 | 5.14 | 5.92 | 5.65 |
| Exp.[47] | | | | 5.9 |

## 4. Computed properties of bulk $MoO_3$

The choice of the *U* parameter to be used for the study of the $MoO_3$ bulk phase study turned out to be more complex than for the $V_2O_5$ case. The first DFT+*U* (*U* = 6.3 eV) calculation reported by Coquet and Willock[48] on $MoO_3$ did not include any dispersion correction but nevertheless led to quit good lattice parameters for the bulk phase, and to a calculated band-gap of 2.1 eV ($MoO_3$ has an experimental indirect band gap of 2.2 eV[6]). The formation enthalpy GGA+*U* calculations by Lutfalla et al.[27] resulted in a large $U_{Mo}$ value of 8.6 eV for Mo(*d*) and a *c*-parameter largely overestimated (4.5 Å), probably due to the absence of dispersion forces. Akande et al. calculated the lattice parameters of α-MoO$_3$ using $U_{Mo}$ = 4.3 and $U_{Mo}$ = 6.3 eV, obtaining an excellent agreement, while the band-gap resulted largely independent on $U$, being around 2 eV even for large $U$ values. Using the HSE06[49,50] hybrid functional these authors computed a direct band-gap of 3.1 eV, similar to the experimental direct gap of 3.0-3.3 eV,[6,51] obtained from optical absorption measurements.[52] Finally, Inzani et al.[21] have checked many possible formulations of Van der Waals dispersive energy contributions, along with $U$ corrections on Mo($d$). They concluded that even though the geometry was better described with vdW-DF2 approach, almost no effect on the band-gap resulted from a variation of $U$ on Mo($d$) from 2 to 8 eV.

We also studied the properties of MoO$_3$ varying the $U$ parameter on the Mo 4d states from 4.3 to 8.6 eV, but we could only confirm the conclusion that there is no effect on the properties of the material. Therefore, no $U$ value on Mo has been included in the calculations. However, as it will be discussed in more detail below, the description of the Kohn-Sham band gap and of other properties slightly improves if a $U$ value is applied to the O 2p orbitals. Use of $U$ parameters for the more delocalized orbitals of the ligand atoms has been suggested in the past, and it turned out to be of some help for the description of the system.[53] When we introduced the $U$ term on O 2p states, we checked that the O$_2$ molecule's binding energy is properly described, in order to be able to have a computational setup capable to describe also O removal from the material (vacancy formation energy). Based on a series of test calculations, see SI **Section S1** and **S2**, we came to the conclusion that a good description of the O$_2$ dissociation energy, of the lattice constants, of the indirect band-gap, of the formation enthalpy, and of the static dielectric constant of bulk MoO$_3$ can be obtained using $U_O$ = 5 eV on O($p$) states. Further details can be found in the next subsections. Therefore, the calculations reported in the following are based on a PBE+U/D3/SOC approach where $U$ = 5 eV has been applied to the O 2p states, and not to Mo 4d orbitals.

*4.1 Structural parameters*

The crystal structure of α-MoO$_3$ is centro-symmetric with orthorhombic symmetry and space group *Pbnm*. The lattice parameters are *a* = 3.963Å , *b* = 13.860Å and *c* = 3.697Å,[52] **Figure 2**.

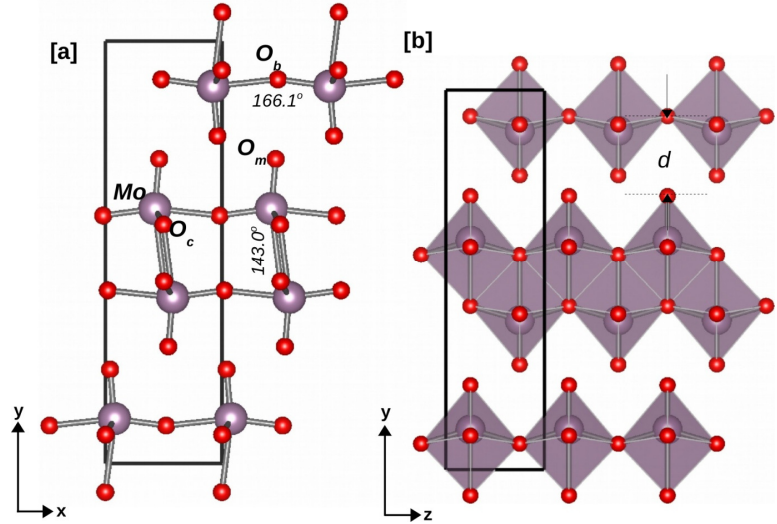

**Figure 2**: Optimized bulk crystal structure of α-MoO$_3$ (PBE+U/D3/SOC functional with $U_O$ = 5.0 eV).

In panel [a] the unit cell is marked with a black solid line and the optimized angles are reported; in panel [b], the MoO$_6$ octahedra forming a chain along the *z*-axis are shown, separated by a zigzag like vdW spacing *d*. The computed structural parameters and band gap values are reported in **Table 3.** We notice that the dispersion corrections are larger than for V$_2$O$_5$, with a reduction of the interlayer spacing by nearly 0.7 Å.

**Table 3**: Optimized bulk structure parameters of MoO$_3$ from PBE and PBE+U/D3/SOC functionals ($U_O$ = 5.0 eV) compared with experimental data. In the parentheses is reported the % error with respect to experiment.

|  | Orthorhombic α -MoO$_3$ (*P bnm*) | | |
| --- | --- | --- | --- |
| Parameters | PBE | PBE+U/D3/SOC | Exp.[6,52] |
| a(Å) | 3.935 | 3.905 (-1.4%) | 3.963 |
| b(Å) | 15.644 | 14.272 (+2.9%) | 13.860 |
| c(Å) | 3.690 | 3.678 (-0.5%) | 3.697 |
| vdW Spacing, *d* (Å) | 3.355 | 2.680 (+1.4%) | 2.644 |
| Mo-O$_m$ (Å) | 1.680 | 1.671 | 1.673 |
| Mo-O$_b$ (Å) | 1.760, 2.198 | 1.751, 2.183 | 1.738, 2.242 |
| Mo-O$_c$ (Å) | 2×1.947 | 2×1.939 | 2×1.948 |
| | 1×2.474 | 1×2.503 | 1×2.316 |
| <Mo-O$_b$-Mo (°) | 167.4° | 166.1° (-1.8%) | 169.2° |
| <Mo-O$_c$-Mo (°) | 142.7° | 143.0° (-0.1%) | 143.2° |
| Band-gap (eV) | 1.99 | 1.88 (-14.5%) | 2.2 |

The results show that the geometry of bulk MoO$_3$ is well reproduced, with typical errors on lattice parameters, buckling angles, and vdW spacing smaller than 3% and usually much smaller. In this respect, the present approach is a bit less accurate than for the case of V$_2$O$_5$, Table 1, where the maximum error is of 1.4%.

*4.2 Band gap and electronic structure*

The bulk MoO$_3$ orthorhombic phase, has indirect band-gap of 2.2 eV as recently validated by experimental absorption spectra and theoretical data from PBEsol-vdW-D2 mehtod.[6] In our PBE+U/D3/SOC calculations with $U_O$ = 5 eV we obtain an indirect band-gap of 1.88 eV, about ~15% lower than the experimental one. On the other hand, the direct band-gap at Γ-point is ~3.0 eV, which is quite close to to the optical absorption measurements which report a value of ~3.0-3.3 eV for this oxide.

The total band structure of the bulk phase is shown on the right side panel of **Figure 4**, which shows the indirect type band gap for this material in pristine phase similar to V$_2$O$_5$. The band gap is due to the indirect transition from R-to-Γ within the high symmetry k-points of the first Brillouin zone of bulk MoO$_3$. For further details see SI **Section 2**, **Table 8**. The total DOS is shown in **Figure 4**. The top panel shows that the empty Mo(*d*) states contribute to the bottom of the conduction band, whereas top of valence band is made of O(*p*) characters, mainly. The split of the Mo(*d*) bands, due crystal field effects, is even larger than in V$_2$O$_5$, reflecting the stronger octahedral field. The calculated O(*p*) valence band width (~6 eV) perfectly matches the known experimental data.[54]

The nature of the chain and bridge oxygens is partly different from that of the molybdenyl oxygen atom. (see also SI **Section S4, Figure 1**).

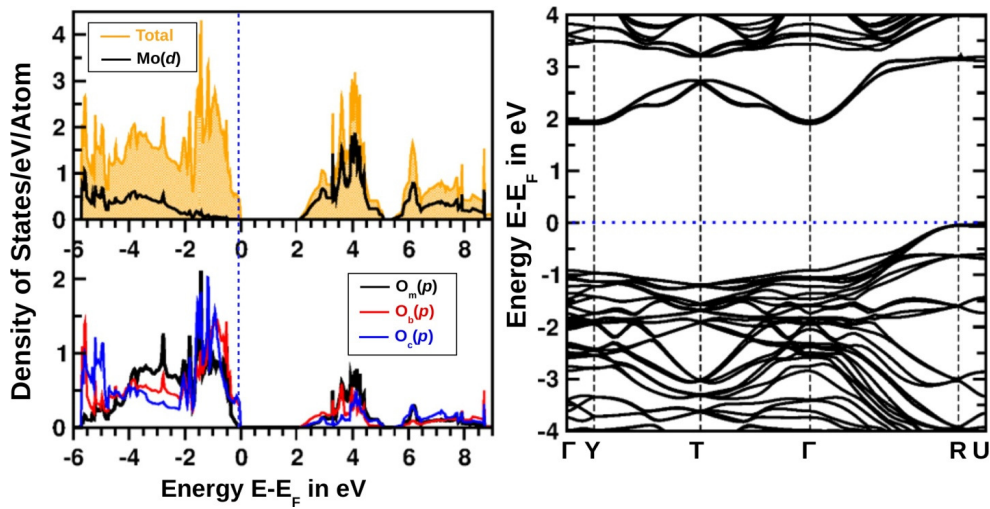

**Figure 4**: Total, Mo(*d*), and O(*p*) projected density of states (left panel) and band-structure (right panel) of bulk $MoO_3$ (PBE+U/D3/SOC). The Fermi level is set to the top of the valence band.

*4.3 Formation Enthalpy*

The formation enthalpy (*ΔH*) of $MoO_3$ computed at the PBE+U/D3/SOC level with respect to the standard references, Mo(s) and $O_2$(g), is 6.77 eV/f.u. (650 KJ/mol). This has to be compared with the experimental value of 7.76 eV (745 KJ/mol). The computed value is underestimated by about 1 eV, or 12.7%. Notice that with the present approach the $O_2$ dissociation energy is 5.53 eV, to be compared with the experimental value of 5.2 eV. Thus, one can consider the present set-up sufficiently accurate, within about 10% error, to describe reactions involving oxygen release from the oxide phase with formation of gas-phase oxygen and a reduced $MoO_{3-x}$ phase. Using a standard PBE approach the computed *ΔH* is 9.36 eV, i.e. 1.6 eV larger than in the experiment, and the $O_2$ dissociation energy, 6.65 eV, is largely overestimated.

*4.4 Static Dielectric Constant*

Finally we consider the dielectric constant of $MoO_3$ as obtained with the PBE+U/D3/SOC ($U_O$ = 5 eV) approach, **Table 4**. The three components of the tensor are reported along with the average value, 5.88. This is quite close to the experimental value, 5.7,[55] and to a calculated value by Lajaunie et al.[56] $\varepsilon_{avg}(0)$ = 5.3 for bulk α-$MoO_3$. Also in this case we do not observe a particular improvement by going from the simple PBE approach to the PBE+U/D3/SOC one: the changes in ε are negligible.

**Table 4**: Calculated static dielectric constant, ε, of bulk MoO$_3$ from PBE and PBE+U/D3/SOC approaches compared with experiment.

| Functional | Orthorhombic α-MoO$_3$ (*Pbnm*) | | | |
|---|---|---|---|---|
| | $\varepsilon_{1xx}(0)$ | $\varepsilon_{1yy}(0)$ | $\varepsilon_{1zz}(0)$ | $\varepsilon_{avg}(0)$ |
| PBE | 6.198 | 5.638 | 5.786 | 5.87 |
| PBE+U/D3/SOC | 6.169 | 5.725 | 5.738 | 5.88 |
| Exp.[55] | | | | 5.7 |

## 5. V$_2$O$_5$ and MoO$_3$ ultrathin layers

### *5.1 V$_2$O$_5$ monolayer*

The monolayer (ML) of V$_2$O$_5$ was cut along the <001> direction. After full optimization of the cell and atomic positions, the geometry was compared to that optimized for the bulk (PBE+U/D3/SOC). The in-plane lattice parameter of the V$_2$O$_5$ ML, *a*, decreases by about 2.6% compared to bulk, whereas the *b* lattice constant remains practically unchanged. This leads to small changes in the bridge angle, from 149.3° to 146.3°. This means that the ML is slightly more buckled after full optimization. This small change has some effect on the width of the ML film (4.394 Å) compared to the experimental bulk phase single layer thickness (4.158 Å).

The calculated DOS of the V$_2$O$_5$ ML is shown in the **Figure 5** where is compared with that of bulk.

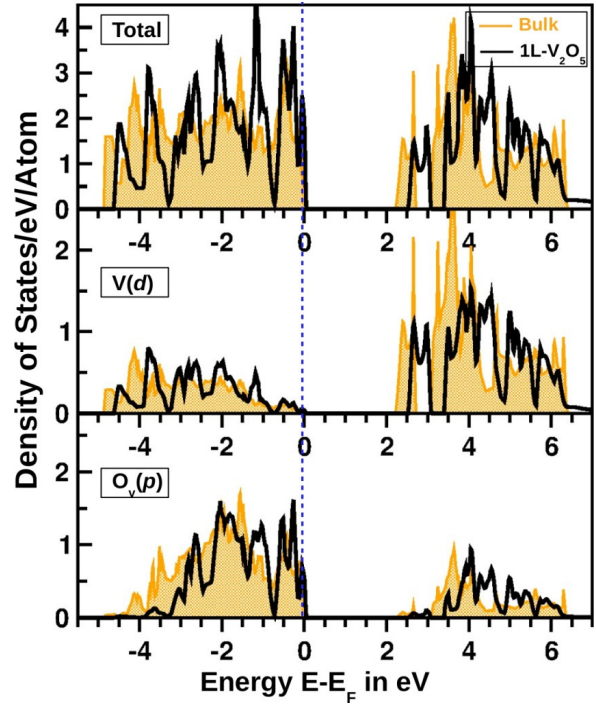

**Figure 5**: Total density of states of $V_2O_5$: top, monolayer of $V_2O_5$ (orange filled) compared to bulk $V_2O_5$ (black solid line); center: projection over V(*d*) states; bottom: projection over $O_v(p)$ states.

There is a clear increase of the band-gap, from 2.18 in bulk $V_2O_5$ to 2.61 eV in $V_2O_5$ ML. In particular, the bottom of the V(*d*) conduction band moves upwards, and the width of the O(*p*) band is reduced by about 0.5 eV, compared to the bulk phase value. The separation of the $d_{xy}$ bands from the rest is preserved for the ML. In general, the calculated band structure of the ML remains similar to the bulk phase, with an indirect band-gap due to the inter band transition from R-to-Γ point of the IBZ. An indirect type band-gap for a $V_2O_5$ ML has been reported also from LDA calculations.[19] For further details see the SI, **Section S5**.

### 5.2 MoO₃ Monolayer

Differently from $V_2O_5$ ML, where a non-negligible contraction of 2.6% was found after the optimization of the lattice parameters, for the case of $MoO_3$ ML the structure (distances and bond angles) remains almost unchanged compared to the bulk (the changes in geometrical parameters are less than 0.5%). This may be due to the very compact $MO_6$ octahedra which are connected via strong covalent bonds. The width of the zigzag ML of $MoO_3$ is 6.377 Å, which is comparable to that of a single layer in the bulk phase, 6.356 Å.

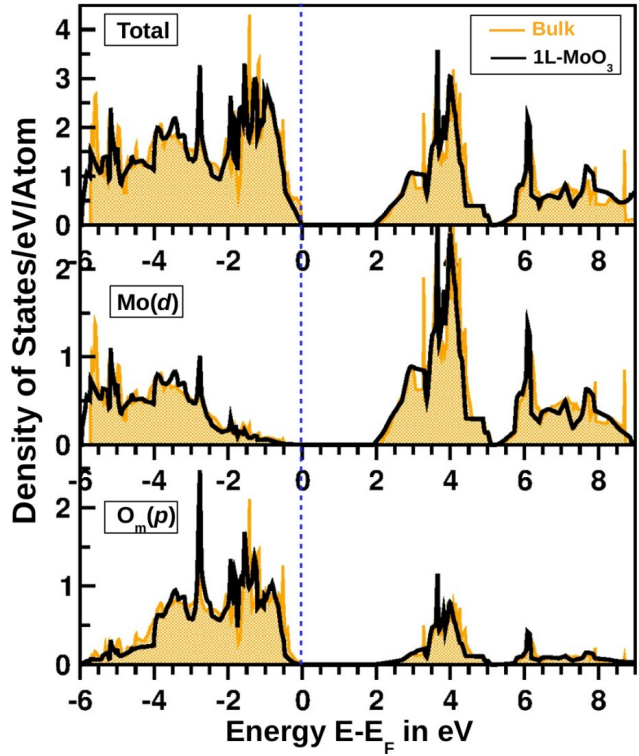

**Figure 6**: Total density of states of $MoO_3$: top, monolayer of $MoO_3$ (orange filled) compared to bulk $MoO_3$ (black solid line); center: projection over V(*d*) states; bottom: projection over $O_v$(*p*) states.

The calculated DOS of a $MoO_3$ ML is shown in **Figure 6**, along with the corresponding DOS for the bulk. The nature of the Mo(*d*) states remains unchanged with similar Mo(*d*) orbitals splitting, but the $d_{xy}$ band component exhibiting a smaller boreìadness in the $MoO_3$ ML. The width of O(*p*) band slightly increases ~6 eV. We notice that the fundamental band-gap also remains indirect (R-to-Γ) and a little increase (60-70 meV) is observed in the band-gap value compared to the bulk. Further details are given in the SI, **Section S5**.

*5.3 Bi- and Tri-layer of $V_2O_5$*

The full optimization of the $V_2O_5$ bilayer film leads to an increase of the in-plane lattice parameter *a* by 0.1 Å compared to the ML, whereas the *b* parameter remains unchanged. The calculated inter-layer space, 2.414 Å, is slightly larger than the bulk experimental value, 2.321 Å. The two pyramidal chains of $VO_5$, projected along the <100> directions in each layer, are inequivalent due their difference in bridging angle by 9° i.e. two layers are buckled. On the contrary, the chain angle remains unchanged (within ~1-2°). The calculated V-O bond lengths are almost coincident with those of bulk $V_2O_5$.

In tri-layer $V_2O_5$ films, the two external layers are very similar to the bi-layer case, whereas the middle layer has a structure which almost coincides with that of bulk $V_2O_5$. The calculated bridge and chain angles in the middle layer are 148.3° and 143.7°, while the calculated bond lengths are V-$O_v$ = 1.600 Å, V-$O_b$ = 2×1.795 Å, V-$O_c$ = 2×1.905, 1×2.023 Å. Comparing these data with those of **Table 1**, we conclude that a $V_2O_5$ tri-layer has a geometric structure which is almost converged to that of the bulk (see supporting data **Section S6** for further details). The calculated DOS of the bi-layer (left panel) and tri-layer (right panel) of $V_2O_5$ are shown in **Figure 7**.

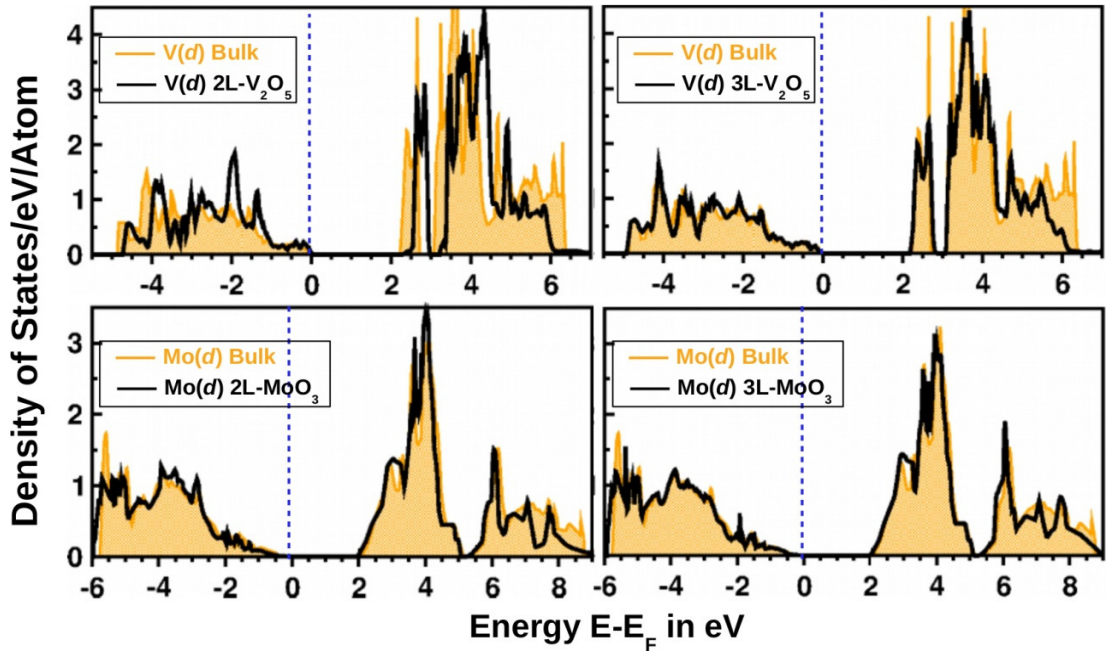

**Figure 7**: Top: Projected V(*d*) DOS of a $V_2O_5$ bi-layer (left panel) and tri-layer (right panel) compared to the bulk. Bottom: Projected Mo(*d*) DOS of a $MoO_3$ bi-layer (left panel) and tri-layer (right panel) compared to the bulk. The Fermi level, $E_F$ is set to the top of the valence band.

The calculated fundamental indirect band-gap for the bi-layer is 2.391 eV, and is further reduced to 2.328 eV in the tri-layer $V_2O_5$, almost converged to the experimental band-gap of the bulk material, 2.30 eV. A comparison of the total V($d$) DOS for the 2L and 3L to the bulk phase is shown in **Figure 7**. As for the geometry, also the electronic DOS of a $V_2O_5$ tri-layer reasonably matches that of the bulk phase.

*5.4 Bi- and Tri-layer of MoO₃*

In a $MoO_3$ bi-layer the vdW spacing is 0.1 Å longer than that calculated for the bulk, $d$ = 2.680Å; the width of the zigzag single layer, 6.361 Å, is close to bulk phase, 6.356 Å. Going to the tri-layer model, the in-plane $a$ and $c$ lattice parameters become almost identical to the experimental bulk phase values, while we observe an additional increase of the vdW spacing by 0.1 Å after full relaxation. The geometrical properties of a $MoO_3$ tri-layer slab of 2.1 nm thickness are thus fully converged to the bulk (SI, **Section S7**).

Also the electronic band structure of the $MoO_3$ bi-layer and tri-layer films is very similar to that of the monolayer. A small reduction of 15-20 meV is found in the indirect band-gap of the bi-layer compared to the bulk, while virtually no change is observed for the three-layer (see SI, **Section S7**). These results further corroborate the conclusions reported in the experimental-theoretical study by Molina-Mandoza et al.[6] where virtually no shift of the band-gap has been observed from the optical spectra of the ML and the bulk phase. This is consistent with the DOS plots of the bi-layer (left panel) and tri-layer (right panel) of $MoO_3$ films, bottom panels of **Figure 7**, which show negligible changes of the Mo($d$) or O($p$) states compared to the bulk phase.

**Conclusions**

In summary, we have studied thermodynamical, electronic properties and calculated dielectric constant of the two well known layered oxides $V_2O_5$ and $MoO_3$ from first-principles PBE-GGA-D3+$U$ calculations in plane-wave with pseudopotential formulation and Gaussian basis hybrid or self consistent hybrid calculations, compromising the computational cost and reasonable accuracy than the known experimental data. The calculated bulk band-gap and formation enthalpy of the both phases were suitable than the known experimental data using this new functionality in our theoretical approach. We comment here with caution that the correlation effect in Mo($d$) is relatively smaller than the V($d$), thus Hubbard $U$ was applied on the O($p$) of bulk MoO$_3$ oxide. Our calculations also confirms the measured band-gap for bulk V$_2$O$_5$ is indirect (2.3 eV) and in case of MoO$_3$ we predicts that the lower band-gap ~2 eV from indirect transition, consistent with existing one theoretical literature, but the higher band-gap ~3.0 eV is from direct transition at Γ-point, which predict further clear experimental measurement to reconfirm our guess. We have carried out extensive study also on their mono and few layers of these oxides and concluded that minimum three layers (1-2 nm width) is needed to get conversed properties like the bulk as soon as the structure and electronic properties are concerned. In case of the V$_2$O$_5$, we expect less possibility of exfoliation due to its stronger vdW forces, than the other oxide MoO$_3$. The impact of the spin-orbit coupling is marginal (less than 10 meV per f.u.) for V$_2$O$_5$ on total energy and about 100 meV per f.u. for MoO$_3$. The point defects for example, oxygen vacancy might have very important role for such mono or few layer geometry, electronic, magnetic and polaronic properties in comparison to same in the bulk structure defects, which could be mater of future research works.

We would like to comment also on the role of different oxygen vacancy for the reduced phase of the MoO$_3$ needs to apply a Hubbard $U$ also over the Mo($d$) orbital, in order to localize two unpaired electrons in the bulk geometry from single vacancy of $O_m$. Whereas such point defects physics and chemistry remains an open issue to the role of polaron (mono or bi?) localization in this particular material in low dimensionality finite size to the bulk geometry, which we would like to address in our future work. More interestingly, the dielectric constant based hybrid approach would another possibility to recheck for these layered oxides, in support to the known experimental data till date about the polaron localization. On the other hand, two unpaired electron from $O_v$ vacancy in bulk V$_2$O$_5$ was possible to localize.

**Acknowledgement**

**References**

[1] Novoselov, K. S.; Geim, A. K. Morozov, S. V.; Jiang, D.; Zhang, Y.; Duboson, S. V.; Griogorieva, I. V.; Firsov, A. A. Electric Field Effect in Atomically Thin Carbon Films. *Science*, **2004**, 306, 666-669.

[2] Geim, A. K. Graphene: Status and Properties. *Science*, **2009**, 324, 1530-1534.

[3] Splendian, A.; Sun, L.; Zhang, Y.; Li, T.; Kim, J.; Chim, C-.Y.; Galli, G.; Yang, F. Emerging Photoluminescence in Monolayer $MoS_2$. *Nano Lett.,* **2010**, 10, 1271-1275.

[4] Chhowalla, M.; Liu, Z.; Zhang, H. Two-dimensional transition metal dichalcogenide (TMD) nanosheets. *Chem. Soc. Rev.*, **2015**, 44, 2584-2586.

[5] Pacchioni, G.; Giordano, L.; Baistrocchi, M. Charging of Metal Atoms on Ultrathin MgO/Mo (001) Films. *Phys. Rev. Lett.*, **2005**, 94, 226104.

[6] Molina-Mendoza, A. J.; Lado, J. L.; Island, J. O.; Niño, M. A.; Aballe, L.; Foerster, M.; Bruno, F. Y.; López-Moreno, A.; Vaquero-Garzon, L.; van der Zant, H. S. J.; Rubio-Bollinger, G.; Agraït, N.; Pérez, E. M.; Fernández-Rossier, J.; Castellanos-Gomez, A. Centimeter-Scale Synthesis of Ultrathin Layered $MoO_3$ by van der Waals Epitaxy. *Chem. Mater.* **2016**, 28, 4042-4051.

[7] Liu, H.; Lee, C. J. J.; Jin, Y.; Yang, J.; Yang, C.; Chi, D. Huge Absorption Edge Blue shifts of Layered-$MoO_3$ Crystals upon Thickness Reduction Approaching 2D Nanosheets. *Phys. Chem. C* **2018**, 122, 12122-12130.

[8] Hanlon, D.; Backes, C.; Higgins, T. M.; Hughes, M.; O'Neill, M.; King, P.; McEvoy, N.; Duesberg, G. S.; Sanchez, B. M.; Pettersson, H.; Nicolosi, V.; Coleman, J. N. Produc-tion of Molybdenum Trioxide Nanosheets by Liquid Exfoliation And Their Applications for High-preformance Supercapacitors. *Chem. Mater.* **2014**, 48, 1751-1763.

[9] Chen, K.; Khodakov, A.; Yang, J.; Bell, A. T.; Iglesia, E. Effect of Catalyst Structure on Oxidative Dehydrogenation of Ethane and Propane on Alumina-Supported Vanadia. *J. Catal.* **1999**, 186, 325.

[10] Alexopoulos, K.; Reyniers, M.-F.; Marin, G. B. Reaction path analysis of propane selective oxidation over $V_2O_5$ and $V_2O_5/TiO_2$. *J. Catal.* **2012**, 289, 127-139.

[11] Rui, X.; Lu, Z.; Yu, H.; Yang, D.; Hng, H. H.; Lim, T. M.; Yan, Q.; Ultrathin $V_2O_5$ nanosheet cathodes: realizing ultrafast reversible lithium storage. *Nanoscale* 2013, 5, 556.

[12] Liu, M.; Su, B.; Tang, Y.; Jiang, X.; Yu, A. Recent Advances in Nanostructured Vanadium Oxides and Composites for Energy Conversion. *Adv. Energy Mater.* **2017**, 7, 1700885(1-34).

[13] Yao, J. N.; Hashimoto, K.; Fujishima, A. Photochromism induced in an electrolytically pretreated $MoO_3$ thin-film by visible light. *Nature* **1992**, 355, 624.

[14] Kitao, M.; Yamada, S.; Hiruta, Y.; Suzuki, N.; Urabe K. Electrochromic absorption spectra modulated by the composition of $WO_3$-$MoO_3$ mixed films. *Appl. Surf. Sci.* **1988**, 33/34, 812.

[15] Pernicone, N.; Lazzerin, F.; Liberti, G.; Lanzavecchia, G. The oxidation of methanol over pure $MoO_3$ catalyst. *J. Catal.* **1969**, 14, 391-393.

[16] Wang, J.; Dong, S.; Yu, C.; Han, X.; Guo, J.; Sun, J. An effcient $MoO_3$ catalyst for in-practical degradation of dye wastewater under room conditions *Catal. Commun.* **2017**, 92, 100-104.

[17] Ferroni, M.; Guidi, V.; Martinelli, G.; Sacerdoti, P.; Nelli, P.; Sberveglieri, G. $MoO_3$-based sputtered thin-films for fast $NO_2$ detection. *Sens. Actuators B* **1998**, 48, 285.

[18] Ranea, V. A.; Qniña, P. L. D. The structure of the bulk and the (001) surface of $V_2O_5$. A DFT+*U* study. *Mater. Res. Express* **2016**, *3*, 085005.

[19] Bhandari, C., Lambrecht, W. R.; Schilfgaarde, M. V. Quasiparticle self-consistent GW calculations of the electronic band structure of bulk and monolayer $V_2O_5$. *Phys. Rev. B* **2015**, 91, 125116.

[20] Li, F.; Chen. Z. Tuning electronic and magnetic properties of $MoO_3$ sheets by cutting, hydrogenation, and external strain: a computational investigation. *Nanoscale* **2013**, 5, 5321.

[21] Inzani, K.; Grande, T.; Vullum-Bruer, F.; Selbach, S. M. A van der Waals Density Functional Study of $MoO_3$ and Its Oxygen Vacancies. *J. Phys. Chem. C* **2016**, 120, 8959-8968.

[22] Dudarev, S. L.; Botton, G. A.; Savrasov, S. Y.; Humphreys, C. J.; Sutton, A. P. Electron-energy-loss spectra and the structural stability of nickel oxide: An LSDA+U study *Phys. Rev. B* **1998**, 57, 1505-1509.

[23] A. D. Becke. Correlation energy of an inhomogeneous electron gas: A coordinate-space model. *J. Chem. Phys.*, **1988**, 88, 1053.

[24] Lee, C.; Yang, W.; Parr, R. G. Development of the Colle-Salvetti correlation-energy formula into a functional of the electron density. *Phys. Rev. B* **1988**, 37(2), 785-789.

[25] A.D. Becke and K.E. Edgecombe. A simple measure of the electron localization in atomic and molecular systems. *J. Chem. Phys.*, **1990**, 92, 5397.

[26] Hu, Z.; Metiu, Z. Choice of *U* for DFT+*U* Calculations for Titanium Dioxides. *J. Phys. Chem. C* **2011**, 112, 5841-5845.

[27] Lutfalla, S.; Shapovalov, V.; Bell, A. T. Calibration of the DFT/GGA+U method for Determination of Reduction Energies of Transition and Rare Earth Metal Oxides of Ti, V, Mo, and Ce. *J. Chem. Theory Comput.* **2011**, 7, 2218-2223.

[28] Skone, J. H.; Govona, M.; Galli, G. Self-consistent hybrid functional for condensed systems. *Phys. Rev. B* **2014**, 89, 195112.

[29] Perdew, J. P.; Burke, K.; Ernzerhof, M. Generalized Gradient Approximation Made Simple. *Phys. Rev. Lett.* **1996**, 77, 3865-3868.

[30] Kresse, G.; Hafner, J. Ab-initio molecular dynamics for liquid metals. *Phys. Rev. B* **1993**, 47, 558-561.

[31] Kresse, G.; Furthmuller, J. Efficient Iterative Schemes for Ab-initio Total-energy Calculations Using a Plane-wave Basis set. *Phys. Rev. B* **1996**, 54, 11169-11186.

[32] Kresse, G.; Joubert, D. From Ultrasoft Pseudopotentials to the Projector Augmented-wave Method. *Phys. Rev. B* **1999**, 59, 1758-1775.

[33] Grimme, S.; Antony, J.; Ehrlich, S.; Krieg, H. A consistent and accurate *ab initio* parameterization of density functional dispersion correction (DFT-D) for 94 elements H-Pu. *J. Chem. Phys.* **2010**, 132, 154104.

[34] Grimme, S.; Ehrlich, S.; Goerigk, J. Effect of the damping function in dispersion corrected density functional theory. *J. Comput. Chem.* **2011**, 32, 1456-1465.

[35] Dovesi, R.; Saunders, V.R.; Roetti, C.; Orlando, R.; Zicovich-Wilson, C. M.; Pascale, F.; Civalleri, B.; Doll, K.; Harrison, N. M.; Bush, I. J.; D'Arco, Ph.; Llunel, M.; Causá, M.; Noël, M.; Maschio, L.; Erba, A.; Rérat, R.; Casassa, S. CRYSTAL17 User's Manual, University of Torino, Torino, Italy, **2018**.

[36] Laun, J.; Oliveira, D.V.; Bredow, T. Consistent gaussian basis sets of double- and triple-zeta valence with polarization quality of the fifth period for solid-state calculations. *J. Comput. Chem.* **2018**, 39(19), 1285-1290.

[37] Kenny, N.; Kannewurf, C. R.; Whitmore, D. H. Optical absorption coeffcients of vanadium pentoxide single crystals. *J. Phys. Chem. Solid.* **1966**, 27, 1237-1246.

[38] Enjalbert, R.; Galy, J. A Refinement of the Structure of $V_2O_5$. *Acta. Cryst. C* **1986**, 42, 1467-1469.

[39] Kohn, W.; Sham, L. J. Self-Consistent Equations Including Exchange and Correlation Effects. *Phys. Rev.* **1965**, 140 (4A), A1133-A1138.

[40] Chiodo, L., García-Lastra, J.M., Iacomino, A., Ossicini, S., Zhao, J., Petek, H. and Rubio, A., 2010. Self-energy and excitonic effects in the electronic and optical properties of TiO 2 crystalline phases. *Physical Review B*, 82(4), p.045207.

[41] Laskowski, Robert, Niels Egede Christensen, Peter Blaha, and Balan Palanivel. "Strong excitonic effects in CuAlO 2 delafossite transparent conductive oxides." *Physical Review B* 79, no. 16 (2009): 165209.

[42] Laubach, S.; Schmidt, P. C.; Thi en, A.; Fernandez-Madrigal, F. J.; Wu, Q-.H.; Jaegermann, W.; Klemm, M.; Horn, S. Theoretical and experimental determination of the electronic structure of $V_2O_5$, reduced $V_2O_{5-x}$ and sodium intercalated $NaV_2O_5$. *Phys. Chem. Chem. Phys.* **2007**, 9, 2564-2576.

[43] Scanlon, D. O.; Walsh, A.; Morgan, B. J.; Watson, G. W. An ab-initio Study of Reduction of $V_2O_5$ through the Formation of Oxygen Vacancies and Li Intercalation. *Phys. Chem. C* **2008**, 112, 9903-9911.

[44] Jovanović. A.; Dobrota, A. S.; Rafailović, L. D.; Mentus, S. V.; Pašti, I. A.; Johansson, B.; Skorodumova, N. V. Structural and electronic properties of $V_2O_5$ and their tuning by doping with 3d elements - modelling using the DFT+U method and dispersion correction. *Phys. Chem. Chem. Phys.* **2018**, 20, 13934-13943.

[45] Gautam, G. S.; Canepa, P.; Abdellahi, A.; Urban, A.; Malik, R.; Ceder, G. The Intercalation Phase Diagram of Mg in $V_2O_5$ from First-Principles. *Chem. Mater.* **2015**, 27, 3733-3742.

[46] Gajdos, M.; Hummer, K.; Kresse, G.; Furthmüller, J.; Bechstedt, F. Linear optical properties in the PAW methodology. *Phys. Rev. B* **2006**, 73, 045112.

[47] Clauws, P.; Vennik, J. Lattice Vibrations of $V_2O_5$. Determination of TO and LO Frequencies from Infrared Reflection and Transmission. *Phys. Status Solidi (b)* **1976**, 76, 707-713.

[48] Coquet, R.; Willock, D. J. The (010) surface of α-$MoO_3$, a DFT+U study. *Phys. Chem. Chem. Phys.* **2005**, 7, 3819-3828.

[49] Heyd, J.; Scuseria, G. E.; Ernzerhof, M. Hybrid Functionals Based on a Screened Coulomb Potential. *J. Chem. Phys.* **2003**, 118, 8207-8215.

[50] Krukau, A. V.; Vydrov, O. A.; Izmaylov, A. F.; Scuseria, G. E. Influence of the Ex-change Screening Parameter on the Performance of Screened Hybrid Functionals. *J. Chem. Phys.* **2006**, 125, 224106.

[51] Akande, S. O.; Chroneos, A; Vasilopoulou, M.; Kennou, S.; Schwingenschlögl, U. Va-cancy formation in $MoO_3$: hybrid density functional theory and photoemission experi-ments. *J. Mater. Chem. C* **2016**, *4*, 9526.

[52] Yin, Z.; Zhang, X.; Cai, Y.; Chen, J.; Wong, J.I.; Tay, Y.Y.; Chai, J.; Wu, J.; Zeng, Z.; Zheng, B.; Yang, H. Y.; Zhang, H. Preparation of $MoS_2$-$MoO_3$ hybrid nanomaterials for light-emitting diodes. *Angew. Chem. Int. Ed.* **2014**, 53, 12560.

[53] Park, S.-G.; Magyari-Köpe, B.; Nishi, Y. Electronic Correlation Effect in Reduced Rutile $TiO_2$ within the LDA+*U* Method. *Phys. Rev. B* **2010**, 82, 11510

[54] Vasilopoulou, M.; Douvas, A. M.; Georgiadou, D. G.; Palilis, L. C.; Kennou, S.; Sygellou, L.; Soultati, A.; Kostis, I.; Papadimitropoulos, G.; Davazoglou, D.; Argitis, P. . The influence of hydrogenation and oxygen vacancies on molybdenum oxides work function and gap states for application in organic optoelectronics. *J. Am. Chem. Soc.* **2012**, 134, 16178-87.

[55] Deb, S. K.; Chopoorian, J. A. Optical Properties and Color Center Formation in Thin Films of Molybdenum Trioxide. *J. Appl. Phys.* **1966**, 37, 4818.

[56] Lajaunie, L.; Boucher, F; Dessapt, R.; Moreau, P. Strong anisotropic influence of local-field effects on the dielectric response of α-$MoO_3$. *Phys. Rev. B* **2013**, 88, 115141.